\documentclass[aps,showpacs,preprintnumbers,amsmath, amssymb]{revtex4}

\usepackage{float}
\usepackage{graphics,epsfig}
\usepackage{graphicx}
\usepackage{dcolumn}
\usepackage{bm}

\begin{document}
\baselineskip=0.8 cm

\title{{\bf No long hair behaviors of ultra-compact objects}}
\author{Guohua Liu$^{1}$\footnote{liuguohua1234@163.com}}
\author{Yan Peng$^{2}$\footnote{yanpengphy@163.com}}
\affiliation{\\$^{1}$ School of Physics and Physical Engineering, Qufu Normal University, Qufu, Shandong, 273165, China}
\affiliation{\\$^{2}$ School of Mathematical Sciences, Qufu Normal University, Qufu, Shandong 273165, China}

\vspace*{0.2cm}
\begin{abstract}
\baselineskip=0.6 cm
\begin{center}
{\bf Abstract}
\end{center}

We investigate distributions of matter fields outside
spherically symmetric ultra-compact objects in the asymptotically
flat background. Based on the dominant energy condition
and the non-negative trace condition, we analytically
find a no long hair behavior, which states that the
effective radius of matter field hairs cannot
extend beyond the outermost null circular orbit.

\end{abstract}

\maketitle
\newpage
\vspace*{0.2cm}

\section{Introduction}

According to the classical general relativity, a characteristic property
of highly curved spacetimes is the existence of null
circular orbits outside ultra-compact objects,
such as black holes and horizonless ultra-compact stars \cite{c1,c2}.
The null circular orbits provide valuable information on the
properties of the spacetime, such as the gravitational
lensing, the black hole shadow and the gravitational waves.
Hence, the null circular orbits have attracted a lot of
attentions from physicists and mathematicians \cite{ad1}-\cite{r2}.

The uniqueness theorem states that black holes are completely characterized
by only three parameters: mass, electric charge and angular momentum,
which are all conserved quantities defined at the
infinity associated with matter fields subject to a Gauss law \cite{ssc1,ssc2,ssc3}.
In contrast, other exterior matter fields not related to a Gauss law
are usually called black hole hairs.
The null circular orbit divides matter field hairs into two parts:
hairs below the orbit and hairs above the orbit.
It is supposed that fields below the orbit tending to fall
into the horizon and fields above the orbit tending to radiate away to infinity.
The authors in \cite{s1} have proposed a physical picture
that self-interaction between these two parts bind
together fields leading to the existence of black hole matter field hairs.
So the null circular orbits may be useful in
describing the distributions of matter field hairs outside black holes.
In accordance with this picture,
the authors in \cite{s1,s2} proposed a no
short hair theorem that the effective radius of matter field hairs must
extend beyond the innermost null circular orbit of spherically symmetric hairy black holes.
In the horizonless ultra-compact star background, such lower bound on the effective radius of hairs
also exists on the non-negative trace condition \cite{YPG}.
Interestingly, it was also found that over half of the matter field hair gravitational mass
is contained above the null circular orbit of spherically symmetric hairy black holes \cite{s2,s3,s4,s5}.
As a further step of lower bound analysis in \cite{s1,s2,YPG}, it is meaningful to examine whether
there is any upper bound on the effective radius of matter field hairs.

This paper is to study exterior matter field hairs outside ultra-compact objects.
The null circular orbits are useful in
describing the distributions of hairs.
We analytically find a no long hair behavior that
the effective radius of matter field hairs must not
extend beyond the outermost null circular orbit.

\section{Investigations on effective radius of matter field hairs}

We consider the gravity system with spherically symmetric
ultra-compact objects surrounded by exterior matter field hairs.
The curved spacetime is characterized by the line element
 \cite{s1,s2}
\begin{eqnarray}\label{AdSBH}
ds^{2}&=&-g(r)e^{-2\chi(r)}dt^{2}+\frac{dr^{2}}{g(r)}+r^{2}(d\theta^2+sin^{2}\theta d\phi^{2}),
\end{eqnarray}
where $\chi(r)$ and $g(r)$ are metric functions depending on the radial coordinate r.
In the asymptotically flat background, the metric functions satisfy
$\chi(r\rightarrow \infty)=0$ and $g(r\rightarrow \infty)=1$ at the infinity.

The Einstein equations $G^{\mu}_{\nu}=8\pi T^{\mu}_{\nu}$ yield
the differential equations
\begin{eqnarray}\label{BHg}
g'=-8\pi r \rho+\frac{1-g}{r},
\end{eqnarray}
\begin{eqnarray}\label{BHg}
\chi'=\frac{-4\pi r (\rho+p)}{g},
\end{eqnarray}
where $\rho$, $p$ and $p_{\tau}$ are interpreted as
the energy density $\rho=-T^{t}_{t}$,
the radial pressure $p=T^{r}_{r}$ and
the tangential pressure $p_{\tau}=T^{\theta}_{\theta}=T^{\phi}_{\phi}$
respectively \cite{s2}. In this work, we take the dominant energy condition
that the pressures are bounded by the non-negative energy density as \cite{s1,s2}
\begin{eqnarray}\label{BHg}
\rho> |p|,~|p_{\tau}|\geqslant 0.
\end{eqnarray}
In this paper, we only focus on the region above the outermost
null circular orbit and it was proved that exterior
null circular orbit outside black hole horizons always exists \cite{er1,er2,er3,er4,er5}. So we can take the condition (4)
without contradiction with the fact that there is $\rho=-p$ on black hole horizons \cite{s1,s2}.

We also take the non-negative trace condition \cite{b1,b2,b3,b4}
\begin{eqnarray}\label{BHg}
T=-\rho+p+2p_{\tau}\geqslant 0.
\end{eqnarray}
An example satisfying the relation (5)
is the gravitating Einstein-Yang-Mills solitons with $T=0$ \cite{T}.
We point out that the non-positive trace condition $T\leqslant 0$ is usually imposed
in the curved spacetimes \cite{s1,s2}.

The metric function $g(r)$ can be putted in the form \cite{b2}
\begin{eqnarray}\label{BHg}
g=1-\frac{2m(r)}{r},
\end{eqnarray}
where $m(r)$ is the gravitational mass contained within a sphere of radial radius r.

With the equation (2), one can further express the mass term by the integral relation
\begin{eqnarray}\label{AdSBH}
m(r)=\int_{0}^{r}4\pi r'^{2}\rho(r')dr'.
\end{eqnarray}

In the following, we deduce the null circular orbit equations \cite{s2}.
The energy and angular momentum are conserved on the geodesic trajectories
since the metric is independent of $t$ and $\phi$.
And the null circular orbit equations can be
obtained from the effective potential
\begin{eqnarray}\label{BHg}
V_{r}=(1-e^{2\chi})E^{2}+g\frac{L^2}{r^2}
\end{eqnarray}
along with the characteristic relations
\begin{eqnarray}\label{BHg}
V_{r}=E^{2}~~~~~~and~~~~~~V_{r}'=0,
\end{eqnarray}
where E and L are conserved energy and
conserved angular momentum respectively.

Substituting equations (2) and (3) into (8) and (9),
we get the null circular orbit equation
\begin{eqnarray}\label{BHg}
N(r)=3g(r)-1-8\pi (r)^2p(r)=0,
\end{eqnarray}
where the discrete roots $r_{\gamma}$ satisfying $N(r_{\gamma})=0$ are the radii of the null circular orbits.

From the expression (7), the finiteness of the gravitational mass implies
\begin{eqnarray}\label{BHg}
r^{2}\rho(r)\rightarrow 0~~~~~~as~~~~~~r\rightarrow\infty.
\end{eqnarray}

Since $p$ is bounded by $\rho$, we obtain the asymptotical infinity behavior
\begin{eqnarray}\label{BHg}
r^{2}p(r)\rightarrow 0~~~~~~as~~~~~~r\rightarrow\infty.
\end{eqnarray}

According to (12), the null circular orbit characteristic equation asymptotically behaves
\begin{eqnarray}\label{BHg}
N(r)\rightarrow 2~~~~~~as~~~~~~r\rightarrow\infty.
\end{eqnarray}

We define $r_{\gamma}^{out}$ as the outermost null circular orbit radius,
which corresponds to the largest positive root of $N(r)=0$.
Above the outermost null circular orbit, $N(r)$ is non-negative as
\begin{eqnarray}\label{BHg}
N(r)>0~~~~for~~~~r\in (r_{\gamma}^{out},\infty).
\end{eqnarray}

The conservation equation $T^{\mu}_{\nu;\mu}=0$ has only one nontrivial component
\begin{eqnarray}\label{BHg}
T^{\mu}_{r;\mu}=0.
\end{eqnarray}

Substituting equations (2) and (3) into (15),
we obtain the pressure equation
\begin{eqnarray}\label{BHg}
p'(r)=\frac{1}{2rg}[(3g-1-8\pi r^2p)(\rho+p)+2gT-8gp],
\end{eqnarray}
where $T=-\rho+p+2p_{\tau}$ is the trace of the energy momentum tensor.

We introduce a new pressure function $P(r)=r^4p$,
which is proved to be useful in describing
the distributions of matter field hairs.
Then the equation (16) can be expressed as
\begin{eqnarray}\label{BHg}
P'(r)=\frac{r}{2g}[N(\rho+p)+2gT]
\end{eqnarray}
with $N=3g-1-8\pi r^2p$.

From (4), (5), (14) and (17), $P(r)$ is an increasing function
above outermost null circular orbits satisfying
\begin{eqnarray}\label{BHg}
P'(r)> 0~~~~for~~~~r\in (r_{\gamma}^{out},\infty).
\end{eqnarray}

We take the condition that the energy density $\rho$ approaches zero faster than
$r^{-4}$, which is imposed to make sure that there are no extra conserved charges besides
the ADM mass (the energy density of Einstein-Maxwell fields associated with electric charges
decreases as the speed of $r^{-4}$ at the infinity) defined at the infinity associated
with matter fields in the spherically symmetric spacetimes \cite{s1,s2}.
Also considering the density dominant condition (4),
the effective pressure satisfies the asymptotical infinity behavior
\begin{eqnarray}\label{BHg}
P(r\rightarrow \infty)=0.
\end{eqnarray}

It was found that the effective pressure $|P(r)|$ must have a local maximum value
at some extremum point $r_{0}$  \cite{s1,s2}. We can define $r_{m}=r_{0}$ as the effective radii
of matter fields. According to (18) and (19), $P(r)$ is an increasing function of r above
the outermost null circular orbit and approaches zero at the infinity.
So the effective radii $r_{m}=r_{0}$ must have an upper bound
\begin{eqnarray}\label{BHg}
r_{m}\leqslant r_{\gamma}^{out}.
\end{eqnarray}

\section{Conclusions}

We investigated distributions of matter fields
outside spherically symmetric ultra-compact objects.
We assumed the dominant energy condition and the non-negative trace condition.
We defined an effective matter field radius at an extremum point,
where the pressure function $|P(r)|$ possesses a local maximum value.
Using analytical methods, we obtained an upper bound
on the effective matter field radius expressed as
$r_{m}\leqslant r_{\gamma}^{out}$ with
$r_{m}$ as the effective matter field radius
and $r_{\gamma}^{out}$ corresponding to the outermost null circular orbit radius.
So we found a no long hair behavior that the
effective matter field radius must not extend beyond the
outermost null circular orbit.

\begin{acknowledgments}

This work was supported by the Shandong Provincial Natural Science Foundation of China under Grant
No. ZR2022MA074. This work was also supported by a grant from Qufu Normal University
of China under Grant No. xkjjc201906, the Youth Innovations and Talents Project of Shandong
Provincial Colleges and Universities (Grant no. 201909118), Taishan Scholar Project of Shandong Province (Grant No.tsqn202103062)
and the Higher Educational Youth Innovation Science
and Technology Program Shandong Province (Grant No. 2020KJJ004).

\end{acknowledgments}


\begin{thebibliography}{99}





\bibitem{c1}
J. M. Bardeen, W. H. Press and S. A. Teukolsky, Rotating black holes: Locally nonrotating frames, energy extraction,
and scalar synchrotron radiation, Astrophys. J. 178,347(1972).

\bibitem{c2}
S. Chandrasekhar, The Mathematical Theory of Black Holes, (Oxford University Press, New
York, 1983).




\bibitem{ad1}
Goebel, C. J.,Comments on the ``vibrations" of a Black Hole,Astrophysical Journal, vol. 172, p.L 95.

\bibitem{ad2}
Teo, E., Spherical Photon Orbits Around a Kerr Black Hole,
General Relativity and Gravitation (2003) 35: 1909.


\bibitem{ad3}
Pedro V.P. Cunha, Carlos A.R. Herdeiro, Eugen Radu,
Fundamental photon orbits: black hole shadows and spacetime instabilities,
Phys. Rev. D 96(2017)no.2,024039.


\bibitem{ad4}
Jai Grover, Alexander Wittig, Black Hole Shadows and Invariant
Phase Space Structures, Phys. Rev. D 96(2017)no.2,024045.

\bibitem{ad5}
Pedro V.P. Cunha, Carlos A.R. Herdeiro, Maria J. Rodriguez,
Does the black hole shadow probe the event horizon geometry?,
Phys. Rev. D 97(2018)no.8,084020.























\bibitem{c3}
S. L. Shapiro and S. A. Teukolsky, Black holes, white dwarfs, and neutron stars: The physics
of compact objects, New York, USA: Wiley(1983)645p.

\bibitem{c4}
V. Cardoso, A. S. Miranda, E. Berti, H. Witek and V. T. Zanchin, Geodesic stability, Lyapunov exponents and quasinormal modes,
Phys. Rev. D 79, 064016(2009).



\bibitem{c5}
S. Hod, Spherical null geodesics of rotating Kerr black holes, Phys. Lett. B 718,1552(2013).





\bibitem{z1}
Emanuel Gallo, J. R. Villanueva, Photon spheres in Einstein and Einstein-Gauss-Bonnet theories and circular null
geodesics in axially-symmetric spacetimes, Phys. Rev. D 92(2015) no.6,064048.

\bibitem{z2}
Zdenek Stuchlik, Jan Schee, Bobir Toshmatov, Jan Hladik, Jan Novotny, Gravitational instability of polytropic
spheres containing region of trapped null geodesics: a possible explanation of central supermassive black holes in
galactic halos, JCAP 1706(2017)no.06, 056.

\bibitem{z3}
Zdenek Stuchlik, Stanislav Hledik, Jan Novotny, General relativistic polytropes with a repulsive cosmological
constant, Phys. Rev. D 94(2016)no.10,103513.





















\bibitem{c6}
Ivan Zh. Stefanov, Stoytcho S. Yazadjiev, Galin G. Gyulchev,
Connection between Black-Hole Quasinormal Modes and Lensing in the Strong Deflection Limit,
Phys. Rev. Lett. 104(2010)251103.





\bibitem{s7}
S. Hod, The fastest way to circle a black hole, Physical Review D 84, 104024 (2011).





\bibitem{s8}
Yan Peng, The extreme orbital period in scalar hairy kerr black holes, Physics Letters B 792(2019)1-3.







\bibitem{us1}
V. Cardoso, A.S. Miranda, E. Berti, H. Witek, and V.T.
Zanchin, Geodesic stability, Lyapunov exponents and quasinormal modes, Phys. Rev. D 79, 064016 (2009).






\bibitem{us2}
S. Hod, Upper bound on the radii of black-hole photonspheres,
Phys. Lett. B 727, 345 (2013).


	




\bibitem{us3}
P. V. P. Cunha, E. Berti, and C. A. R. Herdeiro, Light-Ring Stability for Ultracompact Objects,
Phys. Rev. Lett. 119 (2017)251102.





\bibitem{us4}
Yan Peng, On instabilities of scalar hairy regular compact reflecting stars,JHEP 1810(2018)185.









\bibitem{r1}
Bahram Mashhoon, Stability of charged rotating black holes in the eikonal approximation, Phys. Rev. D 31(1985)no.2,290-293.


\bibitem{r2}
S. Hod,Universal Bound on Dynamical Relaxation Times and Black-Hole Quasinormal Ringing,Phys. Rev. D 75(2007)064013.





\bibitem{ssc1}
R. Ruffini and J. A. Wheeler, Physics Today 24, 30 (1971).

\bibitem{ssc2}
W. Israel, Phys. Rev. 164, 1776 (1967) and Commun. Math. Phys. 8, 245
(1968).

\bibitem{ssc3}
B. Carter, Phys. Rev. Letters 26, 331 (1971); R. M. Wald, Phys. Rev. Letters
26, 1653 (1971).





































\bibitem{s1}
D. N$\acute{u}\tilde{n}$ez, H. Quevedo, and D. Sudarsky,Black Holes Have No Short Hair, Phys. Rev. Lett.
76, 571(1996).






\bibitem{s2}
S. Hod,Hairy Black Holes and Null Circular Geodesics, Phys. Rev. D 84, 124030 (2011).





\bibitem{YPG}
Yan Peng, No short hair behaviors of ultra-compact stars,Eur.Phys.J.C 81(2021)3,245.



\bibitem{s3}
Yun Soo Myung, Taeyoon Moon,Hairy mass bound in the Einstein-Born-Infeld black hole,
Phys. Rev. D 86.084047.


\bibitem{s4}
Yan Peng, Hair mass bound in the black hole with nonzero cosmological constants,
Phys. Rev. D 98(2018)104041.

\bibitem{s5}
Yan Peng, Hair distributions in noncommutative Einstein-Born-Infeld black holes, Nucl.Phys.B 941(2019)1-10.








\bibitem{er1}
S. Hod, Phys. Lett. B 727,345(2013). arXiv:1701.06587.

\bibitem{er2}
P.V.P.Cunha, C.A.R.Herdeiro, Phys. Rev. Lett.124,181101(2020).

\bibitem{er3}
Shahar Hod, Eur. Phys. J. C 82(2022)663.


\bibitem{er4}
S. Hod, Extremal black holes have external light rings, Phys. Rev. D 107,024028(2023), arXiv:2211.15983.


\bibitem{er5}
Yan Peng, The existence of null circular geodesics outside extremal spherically symmetric asymptotically flat hairy black holes, arXiv:2211.14463 [gr-qc].







\bibitem{b1}
S. Hod, Lower bound on the compactness of isotropic ultra-compact objects,
Physical Review D 97, 084018(2018).



\bibitem{b2}
S. Hod, Self-gravitating field configurations: The role of the energy-momentum trace, Physics Letters B 739(2014)383.



\bibitem{b3}
Jan Novotn\'{y}, Jan Hlad\'{i}k, Zden\u{e}k Stuchl\'{i}k,
Polytropic spheres containing regions of trapped null geodesics,
Physical Review D 95(2017)043009.



\bibitem{b4}
S. Hod, Analytic study of self-gravitating polytropic spheres with light rings,
The European Physical Journal C 78(2018)417.



\bibitem{T}
R. Bartnik and J. McKinnon, Particlelike Solutions of the Einstein-Yang-Mills Equations,
Phys. Rev. Lett. 61(1998)141.




\end{thebibliography}
\end{document}